%% file: Main.tex
\documentclass[sigconf,screen]{acmart}

\usepackage{algorithmic}
\usepackage{graphicx}
\usepackage{textcomp}
\usepackage{xcolor}
\usepackage{enumitem}
\usepackage{hyperref}
\usepackage{comment}
\usepackage{svg}
\usepackage{lipsum}
\usepackage{mdframed}
\usepackage{amsmath,amsfonts}

\AtBeginDocument{%
  \providecommand\BibTeX{{%
    \normalfont B\kern-0.5em{\scshape i\kern-0.25em b}\kern-0.8em\TeX}}}
\setcopyright{acmlicensed}
\acmPrice{15.00}
\acmDOI{10.1145/3611643.3613084}
\acmYear{2023}
\copyrightyear{2023}
\acmSubmissionID{fse23ivr-p20-p}
\acmISBN{979-8-4007-0327-0/23/12}
\acmConference[ESEC/FSE '23]{Proceedings of the 31st ACM Joint European Software Engineering Conference and Symposium on the Foundations of Software Engineering}{December 3--9, 2023}{San Francisco, CA, USA}
\acmBooktitle{Proceedings of the 31st ACM Joint European Software Engineering Conference and Symposium on the Foundations of Software Engineering (ESEC/FSE '23), December 3--9, 2023, San Francisco, CA, USA}
\received{2023-05-03}
\received[accepted]{2023-07-19}

\begin{document}

\title{Towards Understanding Emotions in Informal Developer Interactions: A Gitter Chat Study}

\author{Amirali Sajadi}
\affiliation{%
  \institution{Drexel University}
  \city{Philadelphia, PA}
  \country{USA}}
\email{amirali.sajadi@drexel.edu}

\author{Kostadin Damevski}
\affiliation{%
  \institution{Virginia Commonwealth University}
  \city{Richmond, Virginia}
  \country{USA}}
\email{kdamevski@vcu.edu}

\author{Preetha Chatterjee}
\affiliation{%
  \institution{Drexel University}
  \city{Philadelphia, PA}
  \country{USA}}
\email{preetha.chatterjee@drexel.edu}

\begin{abstract}
Emotions play a significant role in teamwork and collaborative activities like software development. While researchers have analyzed developer emotions in various software artifacts (e.g., issues, pull requests), few studies have focused on understanding the broad spectrum of emotions expressed in chats. As one of the most widely used means of communication, chats contain valuable information in the form of informal conversations, such as negative perspectives about adopting a tool. In this paper, we present a dataset of developer chat messages manually annotated with a wide range of emotion labels (and sub-labels), and analyze the type of information present in those messages. We also investigate the unique signals of emotions specific to chats and 
distinguish them from other forms of software communication. 
Our findings suggest that chats have fewer expressions of \textit{Approval} and \textit{Fear} but more expressions of \textit{Curiosity} compared to GitHub comments. We also notice that \textit{Confusion} is frequently observed when discussing programming-related information such as unexpected software behavior. Overall, our study highlights the potential of mining emotions in developer chats for supporting software maintenance and evolution tools.

\end{abstract}



\maketitle

\input{Introduction.tex}

\input{Methodology.tex}
\input{Results.tex}
\input{Conclusion_Future_Work.tex}

\bibliographystyle{ACM-Reference-Format}
\bibliography{citation, preethabib}

\end{document}

%% file: Introduction.tex
\section{Introduction}
Emotions can greatly influence teamwork and collaborative activities such as software development. Specific tasks have been found to be significantly impacted by developer emotions, e.g., bug fixing \cite{Destefanis2016SoftwareDD, ortu2016}, and build success of continuous integration \cite{7962396}. 
Researchers have extensively studied how developer emotions affect software development, created approaches for automatically detecting emotions \cite{MarValous, deva, chen2019sentimoji, emotxt} and, in cases, provided recommendations for the developers \cite{Ebert2021AnES, Chatterjee21, Fucci21}. More recently, researchers have studied complex, emotionally charged psychological concepts such as toxicity in issue reports~\cite{miller2022did, heated_discussions}, and confusion in code reviews~\cite{Ebert2021AnES}. Likewise affective trust between developers of a project was investigated in pull requests and commit comments~\cite{Sajadi2023, p149, arsenalgsd, calefato2016affective}.

A significant amount of research has been conducted on analyzing the emotions of developers in various software artifacts, such as issues, and pull requests~\cite{esem-e, calefato2016affective, imran2022}. However, surprisingly, 
there is a lack of studies on understanding emotions on chat platforms, despite their widespread use among software developers.
Other aspects of developer chats have been previously studied and it was shown that chats are generally interactive and often used for informal communications \cite{Lin21, chatterjee2021automatically} which intuitively makes these communication platforms a suitable place to express emotions.  
Chatterjee et al.  noticed that expression of developer emotions is prevalent in chat communications 
on platforms such as Slack, IRC, and Discord~\cite{Chatterjee2019, Chatterjee2022Data}. 
Kuutila et al. investigated Slack and Hipchat to analyze developers' sentiments and their impact on productivity~\cite{Kuutila2020ChatAI}. 
However, none of these studies systematically analyze developer emotions in textual chat messages.

In this paper we investigate different types of emotions expressed in developers' chat communications. We first select a subset of 400 developer chat messages from the Gittercom dataset~\cite{gittercom}, and then manually annotate it with emotion categories (e.g., \textit{Anger}, \textit{Joy}) using Imran et al's extended emotion taxonomy~\cite{imran2022}. Additionally, we leveraged Pan et al.’s taxonomy to determine the type of information shared (e.g., programming problems, task progress) in these messages. 
Next, we qualitatively analyze the dataset of 400 messages to  understand the relationship between the type of information conveyed and the type of emotion expressed in the messages. 
We aim to answer the following research questions: \textbf{(RQ1:)} What types of emotions are expressed in developer chats and how are they associated with specific types of information or developer intent?; \textbf{(RQ2:)} How do emotions expressed in chats differ from emotions expressed in other forms of software communications? What are the specific signals of emotions that are unique to chats?

Our findings show that, compared to the GitHub issue or pull request comments, chat messages contain fewer instances of expressing \textit{Approval} and \textit{Fear} and more instances of expressing \textit{Curiosity}, which is expected since chat communications are informal in nature and often follow a Q\&A format. Chat communications also consists of more emoticons, shorter sentences, and a generally more informal tone.
These observations can improve the effectiveness of automatic emotion detection tools by providing insights into the modifications they require for adapting to chat platforms.

%% file: Methodology.tex
\section{Methodology}\label{Dataset}

\noindent
\textbf{Emotion Categories and Detection Tool.}
Shaver's taxonomy, widely used in various software engineering studies~\cite{emtk, esem-e, imran2022}, is a hierarchical, tree-structured emotion representation model, consisting of three levels. The top level comprises six basic  emotions: \textit{Anger}, \textit{Love}, \textit{Fear}, \textit{Joy}, \textit{Sadness}, and \textit{Surprise}. For each basic emotion, there exist secondary and tertiary-level emotions that provide more refined granularity for the preceding level. For instance, \textit{Optimism} and \textit{Hope} are the secondary and tertiary level emotions, respectively, for \textit{Joy}. In a recent work, Imran et al. noticed that certain emotions commonly expressed in developer communications were absent from Shaver's framework~\cite{imran2022}. Therefore, they extended Shaver's categories with select emotions from GoEmotions~\cite{goemotions}. In this study, we use Imran et al's extended taxonomy (Table \ref{tab:shavers_category}). 


\noindent
\textbf{Data Selection.} For this study, we use GitterCom, a dataset, consisting of 10,000 messages collected from 10 Gitter communities~\cite{gittercom}. 
In order to obtain a goldset, we selected a subset of 400 messages to be manually annotated with corresponding emotion labels. In order to obtain a statistically significant sample with confidence of 95\%±5\%, we sampled 400 messages distributing the samples equally across 4 different communities~\cite{bujang2017simplified}.
As a measure to avoid the inclusion of text that does not exhibit any emotion, we have decided to limit our sampling to the instances that contain either a positive or negative sentiment, since intuitively messages with a stronger sentiment have a higher potential for expressing emotion. To preprocess the data, we removed stopwords, urls, and user mentions from the messages. We also converted the text to lower-case, and tokenized the words. NLTK VADER~\cite{vader} was used to automatically assign a value between -1 and 1 to each message, representing its sentiment. The authors then randomly selected 200 instances from the messages with sentiment scores in the first quartile of the value distribution (strongly positive) and 200 instances from the messages with sentiment values in the last quartile of the value distribution (strongly negative). These 400 messages contain 100 chat messages from each of the four GitterCom projects with the highest number of users (i.e., \textit{scikit-learn}, \textit{Marionette}, \textit{jHipster}, and \textit{UIkit}). 

\begin{table}
\centering
\scriptsize
\vspace{2mm}
\caption{Extended Taxonomy of Shaver's Tree-structured Emotion Categories \cite{imran2022}}
\vspace{-3mm}
\begin{tabular} { | p{1cm} | p{1.4cm} | p{4.8cm} | }
\hline
    Basic Emotion & Secondary Emotion & Tertiary Emotion \\ \hline\hline
    
     & Irritation & Annoyance, Agitation, Grumpiness, Aggravation, Grouchiness \\ \cline{2-3}
     & Exasperation & Frustration \\ \cline{2-3}
     & Rage & Anger, Fury, Hate, Dislike, Resentment, Outrage, Wrath, Hostility, Bitterness, Ferocity, Loathing, Scorn, Spite, Vengefulness \\ \cline{2-3}
     Anger & Envy & Jealousy \\ \cline{2-3}
     & Disgust & Revulsion, Contempt, Loathing \\  \cline{2-3}
     & Torment & - \\ \cline{2-3}
     & Disapproval & - \\ \hline
     \hline

     & Affection  & Liking, Caring, Compassion, Fondness, Affection, Love, Attraction, Tenderness, Sentimentality, Adoration \\ \cline{2-3}
    Love & Lust & Desire, Passion, Infatuation \\ \cline{2-3}
     & Longing  & - \\ \hline
    \hline
    & Horror & Alarm, Fright, Panic, Terror, Fear, Hysteria, Shock, Mortification \\ \cline{2-3}
    Fear & Nervousness & Anxiety, Distress, Worry, Uneasiness, Tenseness, Apprehension, Dread \\ \hline
    \hline
     & Cheerfulness & Happiness, Amusement, Satisfaction, Bliss, Gaiety, Glee, Jolliness, Joviality, Joy, Delight, Enjoyment, Gladness, Jubilation, Elation, Ecstasy, Euphoria \\ \cline{2-3}
     & Zest & Enthusiasm, Excitement, Thrill, Zeal, Exhilaration \\ \cline{2-3}
     & Contentment & Pleasure \\ \cline{2-3}
     & Optimism & Eagerness, Hope \\  \cline{2-3}
     Joy & Pride & Triumph \\ \cline{2-3}
     & Enthrallment & Enthrallment, Rapture \\ \cline{2-3}
     & Relief & - \\ \cline{2-3}
     & Approval & - \\ \cline{2-3}
     & Admiration & - \\ \cline{2-3}
     \hline
    \hline
    & Suffering & Hurt, Anguish, Agony \\ \cline{2-3}
    & Sadness & Depression, Sorrow, Despair, Gloom, Hopelessness, Glumness, Unhappiness, Grief, Woe, Misery, Melancholy \\ \cline{2-3}
    & Disappoint & Displeasure, Dismay \\ \cline{2-3}
    Sadness & Shame & Guilt, Regret, Remorse \\ \cline{2-3}
    & Neglect & Embarrassment, Insecurity, Insult, Rejection, Alienation, Isolation, Loneliness, Homesickness, Defeat, Dejection, Humiliation \\ \cline{2-3}
    & Sympathy & Pity \\ \hline 
    \hline
    & Surprise & Amazement, Astonishment \\ \cline{2-3}
    & Confusion & -  \\ \cline{2-3}
    Surprise & Curiosity & -  \\ \cline{2-3}
    & Realization & - \\ \hline 
\end{tabular}
\label{tab:shavers_category}
\vspace{-5mm}
\end{table}


\noindent
\textbf{Dataset Annotation.} Two human judges with (3+ years) experience in programming and familiarity with the Gitter platform annotated the 400 selected messages. The annotators were given instructions, similar to ones presented in Imran et al.'s study~\cite{imran2022}, containing details on emotion categories, subcategories, definitions and examples. The judges were asked to determine whether each message expresses any of the six basic emotions along with their secondary and tertiary subcategories. Since chat messages in the GitterCom dataset are, in many cases, not a single sentence, one can often extract the context for each message exhibiting an emotion. The annotations, therefore, relied on the emotion expressed through the entire chat message rather than single sentences.  

Next, we adopted Pan et al.'s taxonomy to determine the types of information available in the 400 messages~\cite{pan2021automating}. This taxonomy categorizes developers' chat communications into the following categories:
(1) \underline{Problem Report}: Conversation regarding unexpected behaviors or bug reports, containing information about 
    (a) \textit{Programming problems,} (b) \textit{Library Problems}, or (c) \textit{Documentation problems};
(2) \underline{Information Retrieval}: 
    Conversations initiated and carried on with the purpose of acquiring or providing information about a certain topic such as:  
    (a) \textit{Programming information}, (b) \textit{Library information}, (c) \textit{Documentation information}, or (d) \textit{General information} e.g., choice of technology;
(3) \underline{Project Management}: Discussion among contributors and team members about the overall state of their project and the future plans for their work, such as: (a)\textit{Technical discussion} or (b) \textit{Task progress} e.g., release schedules.

Following the initial annotation phase, the inter-rater agreement for each emotion category was calculated using Cohen's Kappa. The resulting values were substantial for \textit{Joy} and \textit{Love} (above 0.6) and moderate for the remaining four emotions (ranging from 0.41 to 0.60) \cite{stemler2004comparison}. To ensure the best possible results, the annotators held multiple discussions to resolve their disagreements and reevaluated their annotations iteratively. This process continued until they reached a Cohen's Kappa value of 1 and resolved all disagreements. 

%% file: Results.tex
\section{Preliminary Results and Discussion}\label{Analysis}

\begin{figure}[]
     \centerline{\includegraphics[width=\linewidth, height=3.9cm]{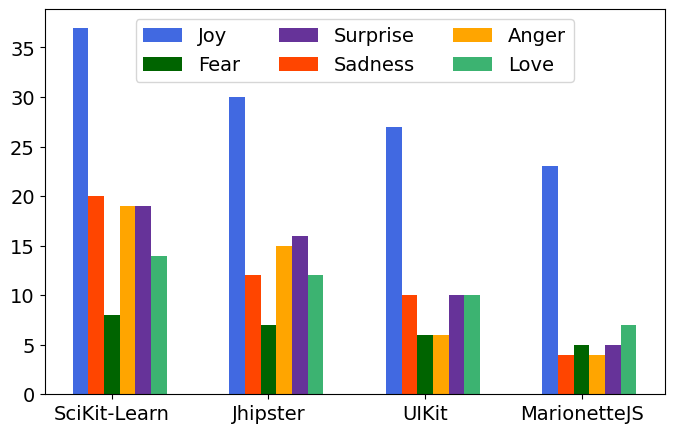}}
    \caption{Project-wise Freq. of Messages Exhibiting Emotion.}
    \label{fig:emotions}
    \vspace{-0.35cm}    
\end{figure}

\vspace{0.2cm}


\noindent

\begin{table*}[]
\vspace{2mm}
    \caption{Example Messages from GitterCom, GitHub Comments, and their Corresponding Annotated Emotions.}
    \centering
    \small
    \begin{tabular}
    {|p{0.15\linewidth}|p{0.38\linewidth}|p{0.38\linewidth}|}
    \hline
    \textbf{Emotion -- \textit{Primary (Secondary, Tertiary)}} & \textbf{Chat Messages} & \textbf{GitHub Comments (Issues and Pull Requests)} \\ \hline
    Anger \textit{(Exasperation, Frustration)}&I can be patient for first time but for each prod build it checks again and thats annoying...& Too bad! Thank you anyway ... This issue is really driving me nuts ...  \\ \hline
    Fear \textit{(Nervousness, Anxiety)}&I always get an error telling me my request can't be processed :worried:&  I can't check this locally because they fail with same error for me even on master  \\ \hline
    Joy \textit{(Cheerfulness, Satisfaction)}&Yea, I'm continually impressed by the community and diversity of the conversation, prs, issues, etc...& You are 100\% on the money with this. Turns out the parsing was incorrect  \\ \hline
    Sadness \textit{(Sadness, Unhappiness)}&@ogrisel what is your plan for the day? I didn't have much time on the weekend unfortunately :-/& I don't think I can, since it is an implementing class. The analyzer is unhappy with it. \\ \hline
    Love \textit{(Affection, Fondness)}&Thank you and i really appreciate your time on this wonderful framework, I love using it :)& PS: I am fan of yours, I love your content out there! :smiley:  \\ \hline
    Surprise \textit{(Confusion, Amazement)}& it did? I didn't see that. haha I know the open tabs issue. Well the ``linear" broke some cases of ``fit'' and ``fit\_transform'' not doing the same thing. Maybe it broke other things, too. 
    & ``true'' feels like magic. Maybe it should be a default value provided to the set or a symbol \\ \hline
    \end{tabular}
    \vspace{-4.5mm}
    \label{examples}
\end{table*}

\begin{mdframed}[backgroundcolor=lightgray!30,topline=false,leftline=false,rightline=false,bottomline=false] 
\noindent
\textbf{RQ1. What types of emotions are expressed in developer chats and how are they associated with specific types of information or developer intent?} 
\end{mdframed}

\noindent
264 out of 400 messages (66\%) contained at least one emotion, while 136 messages (34\%) expressed no emotions. As indicated by Figure \ref{fig:emotions}, \textit{Joy} is the most prevalent emotion expressed in our dataset.  Within the 264 messages that contain emotions, 117 (44.3\%) exhibit \textit{Joy}, 50 (18.9\%) \textit{Surprise}, 46 (17.4\%) \textit{Sadness}, 44 (16.6\%) \textit{Anger}, 43 (16.2\%) \textit{Love}, and 26 (9.8\%) \textit{Fear}. Fig \ref{fig:emotions} illustrates the distribution of the emotions in our dataset across the four Gitter projects. \textit{Joy} is consistently the most frequent emotion across all projects while \textit{Fear} stays the least expressed emotion in three of the projects. Overall \textit{Scikit-Learn}, \textit{Jhipster}, and \textit{UIKit} have rather similar distributions of emotions. MarionetteJS, however, tends to be different since it has a more even distribution over the six emotions.
\begin{figure}[htbp]
  \centering
  \includegraphics[scale=0.2, width=\linewidth, height=.63\linewidth]{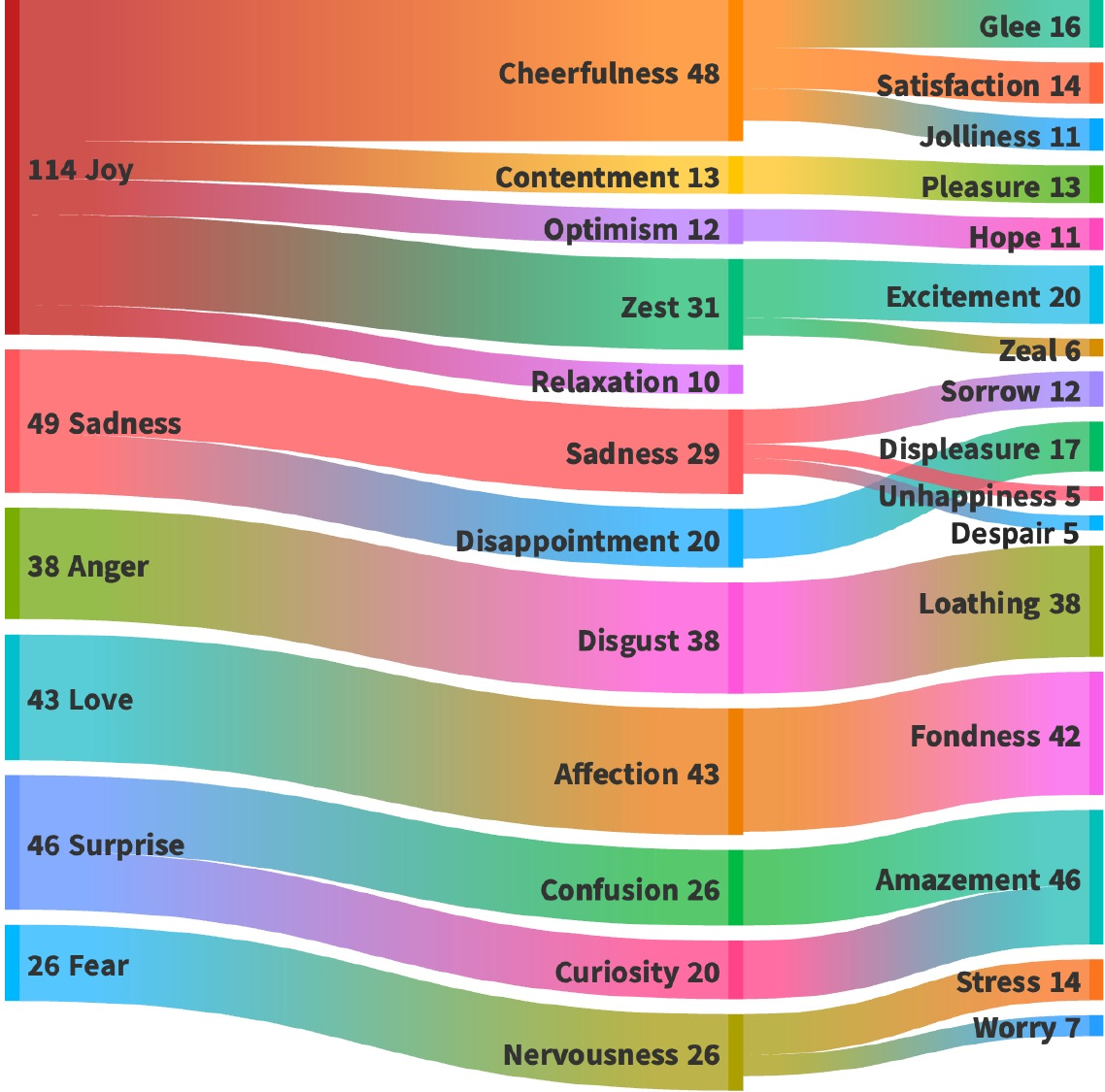}
  \caption{Distribution of the Base, Second, and Third-Level Emotions (n >= 5).}
  \vspace{-7mm}
  \label{sankey}
\end{figure}
In Figure \ref{sankey} we show the results of the annotations for the first, second, and third-level emotions. Overall, \textit{Joy} and \textit{Sadness} stem from more diverse secondary and tertiary categories compared to the other four basic emotions. 
\textit{Joy} is dominated by the second-level emotions of \textit{Cheerfulness} and \textit{Zest}, which can be an indicator of a positive attitude and environment in the communities we explored. \textit{Sadness}, on the other hand, is more evenly split across two of its second-level categories: \textit{Sadness}, which generally expresses a level of dissatisfaction towards the topic, and \textit{Disappointment}, which has often been directed towards unexpected behaviors in software. \textit{Anger} and \textit{Love}, in contrast, are more consistent across the emotion levels. \textit{Love}, for example, almost always is categorized as \textit{Affection} and \textit{Fondness} on the second and third levels.

The annotations of the information categories show that the four most common types of information available in the developers' chats are \textit{Technical progress}, \textit{Programming information}, \textit{Technical discussion}, and \textit{Programming problem} with 31, 30, 28, and 24 instances, respectively. 
The categories related to \textit{Library} and \textit{Documentation} were scarce, with less than 10 occurrences for each category.
These results highlight the prevalence of the exchange of information regarding general project management, programming-related problem reports, and programming-related information retrieval in our dataset.
Furthermore, 
we observed a consistent trend of messages discussing the technical progress of a project being accompanied by positive emotions, such as \textit{Joy} and \textit{Love}.
The messages containing the \textit{Surprise} emotion, are correlated with those inquiring information about programming or those reporting programming-related problems. 
We also noticed that chat conversations frequently involve discussions about unexpected software behavior, which can elicit \textit{Confusion}, a sub-category of \textit{Surprise} in our taxonomy.

\begin{mdframed}[backgroundcolor=lightgray!30,topline=false,leftline=false,rightline=false,bottomline=false] 
\noindent
\textbf{RQ2. How do emotions expressed in chats differ from emotions expressed in other forms of software communications? What are the specific signals of emotions that are unique to chats?}
\end{mdframed}

\noindent
To identify the unique characteristics of developers' chat communications, we compare our emotion annotations on chat messages with the emotions in Imran et al.'s dataset of GitHub issues and pull request comments with either positive or negative sentiment \cite{imran2022}. Table \ref{examples} contains examples of chat messages (from our dataset), along with examples of GitHub comments (from Imran et al.'s dataset) exhibiting the same emotions. 
In general, chat messages tend to facilitate more informal conversations, in keeping with the nature of the communication tool. As one would expect, the length of the messages in chat communications are generally shorter. GitterCom dataset messages are on average 5.87 words long (5.09 in our sampled 400 instances), while the 2000 GitHub comments in Imran et al.'s dataset are on average 12.82 words, i.e., more than twice as long. We also notice a pattern in the more frequent use of emoticons in chat messages compared to GitHub comments. Emoticons are often indicators of implicit emotions, and understanding them could potentially improve the performance of existing emotion detection tools on software engineering-specific text~\cite{chen2019sentimoji}.

We observe that a significant number of instances labeled as \textit{Joy} in the GitHub data ($\sim$18\%) are exhibiting \textit{Approval} on the second level. In contrast, in our Gitter dataset only two instances out of the 117 chat messages that exhibit \textit{Joy} express the writer's approval. This inconsistency is expected since 
GitHub communications often entail the evaluation of one's contribution to the project. For instance, the changes suggested through a pull request in many repositories are required to be approved by at least one reviewer prior to being merged. 
Therefore, approvals exhibiting positive emotions such as \textit{Joy} are more frequent in GitHub.

Developers' chat communications predominantly follow Q\&A formats~\cite{chatterjee2019exploratory}, which may explain the higher prevalence of instances labeled as \textit{Curiosity} in our dataset.
Around 3.5\% of the GitHub issue and pull request comments presented in Imran et al.'s dataset exhibit \textit{Curiosity}, whereas 6.5\% of the Gitter chat communications in our dataset demonstrate some form of this emotion.
GitHub comments may elicit more negative emotions (e.g., \textit{Fear} and by extension its subcategories, \textit{Nervousness}, \textit{Worry}, and \textit{Stress}), since they are often used to discuss issues or bugs in the code and negative emotions such as \textit{Fear} are commonly associated with uncertainty and risk. In line with these expectations, we observe that \textit{Fear} was present in 9.9\% of the GitHub comments and 6.5\% of the chat messages. Imran et al.'s annotations also point to some instances of \textit{Fear} containing \textit{Horror}, a 2nd level emotion absent in the our chat messages. 

%% file: Conclusion_Future_Work.tex
\section{Implications}
To the best of our knowledge, this study took the first step toward systematically analyzing emotions in developer chat communications. 
Our analysis of the Gitter dataset revealed a range of  emotions expressed by developers on chats, predominantly  \textit{Joy}, \textit{Surprise}, and \textit{Sadness}. We noticed that technical progress in software development often evokes positive emotions such as \textit{Joy} and \textit{Love}. In contrast, unexpected software behavior or bugs tends to elicit negative emotions such as \textit{Sadness} and \textit{Anger}. These findings emphasize the potential to develop automated interventions, such as emotion-detection bots,  that take into account users' emotional responses to unexpected events, and suggest potential solutions in a timely manner~\cite{linda2020, 9930282, Lin:2016:WDS:2818052.2869117}. 
In addition to the basic emotions, the secondary and tertiary-level emotions enables us to further analyze the messages and can shed light on the underlying causes of the dominant emotions in informal developer conversations. 
For instance, \textit{Amazement} or \textit{Confusion}, subcategories of \textit{Surprise}, were exhibited in chat messages detailing the satisfactory or unexpected performance of a newly adopted tool.
The dominant presence of \textit{Curiosity} in chat communications compared to the GitHub comments confirms the prevalence of the Q\&A questions in chats. This suggests that chats are better mining source to design Q\&A-based systems such as conversational search assistants~\cite{Chatterjee21, Lin21}.

Limited availability of ground truth data has hindered the extensive evaluation of existing approaches for emotion detection in software developers' written text~\cite{opinionlitreview}. As a first step towards addressing this challenge, we present a dataset of 400 developer chat messages annotated with Imran et al's extended emotion taxonomy~\cite{imran2022}, originally based on Shaver's taxonomy~\cite{shaver1987emotion}. 
Our dataset can be leveraged to analyze developer emotions across various channels (e.g., chats, pull requests) in a software project. Training tools on different communication channels in a project, including chats, offers the potential for building project-specific emotion detectors.

Overall, tools and applications that aim to improve software development processes and team communication can benefit from mining developer chats. Chats can provide rich contextual information on how developers interact and collaborate in real-time, which can help identify communication gaps and improve team dynamics. Proactively identifying negative emotions expressed in chat conversations can help detect potential conflicts, prevent burnout, and improve team collaboration. By analyzing the emotions expressed by developers towards specific aspects of a project, one can assess their opinions on particular tools and technologies.
Compared to other artifacts such as issue comments, chats can offer a more informal and nuanced perspective on developer emotions and interactions. Chats are often more conversational and spontaneous, allowing for a broader range of emotions and expressions (e.g., burnout) that may not be captured in other types of communication.

\section{Conclusion and Future Work}

The increasing reliance on chat platforms for virtual communication among OSS teams and developers in general highlights the significance of studying these communications to gain a better understanding of the development process. As a distinct type of messaging tool, chat platforms provide a unique form of communication that allows developers to express themselves more spontaneously, including exhibiting emotions about specific aspects of a project. 
We present a dataset of 400 messages from Gitter, manually annotated to identify prevalent emotions in developer chats. Our findings shed light on how emotions are expressed on chat platforms and how they differ from emotions expressed in GitHub issue and pull requests. We also explored how the emotions vary based on developer intent and the type of information exchanged in the chat messages. 
Our immediate next steps focus on expanding to a larger dataset, manually annotated with the emotion categories. This step will help us establish, with confidence, the areas in which the current automatic emotion detection tools are lacking (e.g., certain emotions or emotion subcategories) and lay the foundation on which we can further develop and improve these tools. We make our annotated dataset publicly available for future research~\cite{figshare}. 

\vspace{1cm}